\documentclass[superscriptaddress,12pt]{revtex4-1}

\usepackage[pdftex]{graphicx}
\usepackage{graphicx}
\usepackage{dcolumn}%
\usepackage{bm}%
\usepackage{amsmath} 
\usepackage[caption=false]{subfig}
\usepackage{color}

\usepackage{hyperref}
\usepackage{soul}

\usepackage{hyperref}
\usepackage{soul}
\usepackage[pdftex]{graphicx}
\usepackage[caption=false]{subfig}
\begin{document}

\title{Efficient ion acceleration and dense electron-positron plasma creation in  ultra-high intensity laser-solid interactions}

\author{D. Del Sorbo}
\address{York Plasma Institute, Department of Physics, University of York, York YO10 5DD, United Kingdom}
\author{D. R. Blackman}
\address{York Plasma Institute, Department of Physics, University of York, York YO10 5DD, United Kingdom}
\author{R. Capdessus}
\address{Department of Physics SUPA, University of Strathclyde, Glasgow G4 0NG, United Kingdom}
\author{K. Small}
\address{York Plasma Institute, Department of Physics, University of York, York YO10 5DD, United Kingdom}
\author{C. Slade-Lowther}
\address{York Plasma Institute, Department of Physics, University of York, York YO10 5DD, United Kingdom}
\author{W. Luo}
\address{Department of Physics SUPA, University of Strathclyde, Glasgow G4 0NG, United Kingdom}
\address{School of Nuclear Science and Technology, University of South China, Hengyang 421001, China}
\author{M. J. Duff}
\address{Department of Physics SUPA, University of Strathclyde, Glasgow G4 0NG, United Kingdom}
\author{A. P. L. Robinson}
\address{Central Laser Facility, STFC Rutherford-Appleton Laboratory, Oxfordshire OX11 0QX, United Kingdom}
\author{P. McKenna}
\address{Department of Physics SUPA, University of Strathclyde, Glasgow G4 0NG, United Kingdom}
\author{Z.-M. Sheng}
\address{Department of Physics SUPA, University of Strathclyde, Glasgow G4 0NG, United Kingdom}
\address{School of Physics and Astronomy, Shanghai Jiao Tong University, Shanghai 200240, China}
\author{J. Pasley}
\address{York Plasma Institute, Department of Physics, University of York, York YO10 5DD, United Kingdom}
\author{C. P. Ridgers}
\address{York Plasma Institute, Department of Physics, University of York, York YO10 5DD, United Kingdom}


\begin{abstract}
The radiation pressure of next generation ultra-high intensity ($>10^{23}$ W/cm$^{2}$) lasers could efficiently accelerate ions to GeV energies. However, nonlinear quantum-electrodynamic effects play an important role in the interaction of these laser pulses with matter. Here we show that these effects may lead to the production of an extremely dense ($\sim10^{24}$ cm$^{-3}$) pair-plasma which absorbs the laser pulse consequently reducing the accelerated ion energy and laser to ion conversion efficiency by up to 30-50\% \& 50-65\%, respectively. Thus we identify the regimes of laser-matter interaction, where either ions are efficiently accelerated to high energy or dense pair-plasmas are produced as a guide for future experiments.
\end{abstract}

%
%
%
%
%

\maketitle

\section{Introduction}

Ultra-high intensity lasers accelerate ions over much shorter distances than conventional  accelerators (microns compared to many meters) with potential applications in medical physics \cite{caron2015deterministic} as well as in fundamental physics \cite{catani2007next}.  Next generation lasers, such as those comprising the soon to be completed Extreme Light Infrastructure \cite{korn2013eli}, could accelerate ions to GeV energies with 100\% efficiency in principle \cite{capdessus2015influence,esirkepov2004highly}.  However, at the intensities expected to be reached in these laser-matter interactions ($I>10^{23}$ W/cm$^{2}$), the laser very rapidly ionizes the target to form a plasma in which nonlinear quantum-electrodynamic (QED) effects play a crucial role \cite{bell2008possibility,del2017spin,ridgers2017signatures}.  Energetic electrons radiate MeV energy gamma-ray photons by nonlinear Compton scattering. The radiated gamma-ray photons can generate electron-positron pairs in the laser-fields \cite{PhysRev.46.1087} which can radiate further photons.  A cascade of pair production ensues, similar to that thought to occur in extreme astrophysical environments such as pulsar \cite{goldreich1969pulsar} and black hole \cite{blandford1977electromagnetic} magnetospheres.  Pair-plasmas more than eight orders of magnitude denser than currently achievable with  ultra-high intensity lasers could be produced \cite{PhysRevLett.102.105001, sarri2015generation,zhu2016dense}, enabling the study of collective behavior in relativistic pair-plasmas  \cite{chen2015scaling}. \color{black}

 At intensities soon to be reached ($I>10^{22}$ W/cm$^{2}$), the radiation pressure ion acceleration mechanism dominates \cite{macchi2013ion} (with a more favorable scaling of ion energy with laser intensity $\epsilon\propto I$ compared to current experiments in the target normal sheath acceleration regime $\epsilon\propto \sqrt{I}$ \cite{PhysRevLett.84.670,PhysRevLett.84.4108,PhysRevLett.85.2945,wilks2001energetic}). In this acceleration scheme, the electromagnetic momentum carried by the laser pushes forwards the electrons at the front of the target, leaving a charge separation layer and creating an electrostatic field that in turn acts on the ions and leads to their acceleration.  
 There are two regimes of radiation pressure ion acceleration depending on whether the target is thicker or thinner than the relativistic skin depth $\delta_{s}=c/(\sqrt{\gamma_{e}}\omega_{pe})$, where $\gamma_{e}$ is the average Lorentz factor of electrons in the plasma and $c$ the speed of light. $\omega_{pe}=\sqrt{4\pi n_{e^{-}}e^{2}/m_{e}}$ is the electron plasma frequency in which $ n_{e^{-}}$ is the electron density, $e$ is the elementary charge and $m_{e}$ is the electron mass. The case where the target is thicker than $\delta_s$  is known as hole-boring (HB), because the intense radiation pressure of the laser punches a hole in the target, snowploughing ions forwards at an approximately constant speed \cite{robinson2009relativistically,weng2012ultra,tamburini2010radiation}. The case where the target thickness is $\ell\le\delta_s$ \color{black} is known as the light-sail (LS) acceleration \cite{macchi2009light}.  Here the target is sufficiently thin that the ions do not need to snowplough through the undisturbed target and so continuously accelerate \cite{esirkepov2004highly}; the dynamics of the target are then exactly analogous to the LS proposed for spacecraft propulsion \cite{forward1984roundtrip}.  Experiments have been performed in both regimes indicating the expected scaling of ion energy with laser intensity (although complicated by electron heating and the break up of very thin targets) \cite{PhysRevLett.106.014801,badziak2004production,PhysRevLett.100.165002,PhysRevLett.102.095002,macchi2013ion,PhysRevLett.109.185006,PhysRevLett.103.245003,PhysRevLett.108.175005,PhysRevLett.108.225002}.

In this article we use 3D particle-in-cell (PIC) simulations to demonstrate that QED effects can reduce the energy of radiation pressure accelerated ions by up to 50\%.  We show that the key role is played by an electron-positron plasma, created by a pair cascade between the laser and the target. This pair-plasma can reach the relativistically corrected critical density  $n_c^{rel}=\gamma_{e}n_{c}=\gamma_{e}m_{e}\omega_{pe}^{2}/(4\pi e^{2})$, i.e. the density at which its dynamics will strongly affect the propagation of the laser pulse \cite{kaw1970relativistic,palaniyappan2012dynamics}.  In fact this pair-plasma may absorb the laser pulse \cite{nerush2011laser,grismayer2015seeded,grismayer2016laser}.  Consequently, the energy of the accelerated ions is strongly reduced.  Depending on the laser intensity and target density, we identify two regimes of next generation laser-plasma interactions: a regime where laser energy is efficiently converted to pairs and gamma-rays and consequently the opposite regime where laser energy is efficiently converted to ion energy. The identification of the different regimes of  ultra-high intensity laser-solid interaction will be crucial to the choice of parameters for experiments aiming at either ion acceleration or pair-plasma creation.

This article is organized as follows. In Sec.~\ref{section_3D_sims} we demonstrate the quenching of HB ion acceleration due to QED emissions by performing 3D PIC simulations. In Sec.~\ref{section_1D_model} we derive a predictive 1D model for this phenomenon. In Sec.~\ref{section_regimes} we use the derived model to identify the different regimes of quenched HB ion acceleration. In Sec.~\ref{section_light_sail} we extend our analysis to the scheme of LS ion acceleration. A discussion about the phenomenon studied is performed in Sec.~\ref{discussion}. Finally, in Sec.~\ref{Conclusions} we draw conclusions.

\section{Quenching of hole-boring ion acceleration by a self-generated pair-plasma}
\label{section_3D_sims}

\begin{figure}[t]
        \centering
        \includegraphics[width=1.\columnwidth]{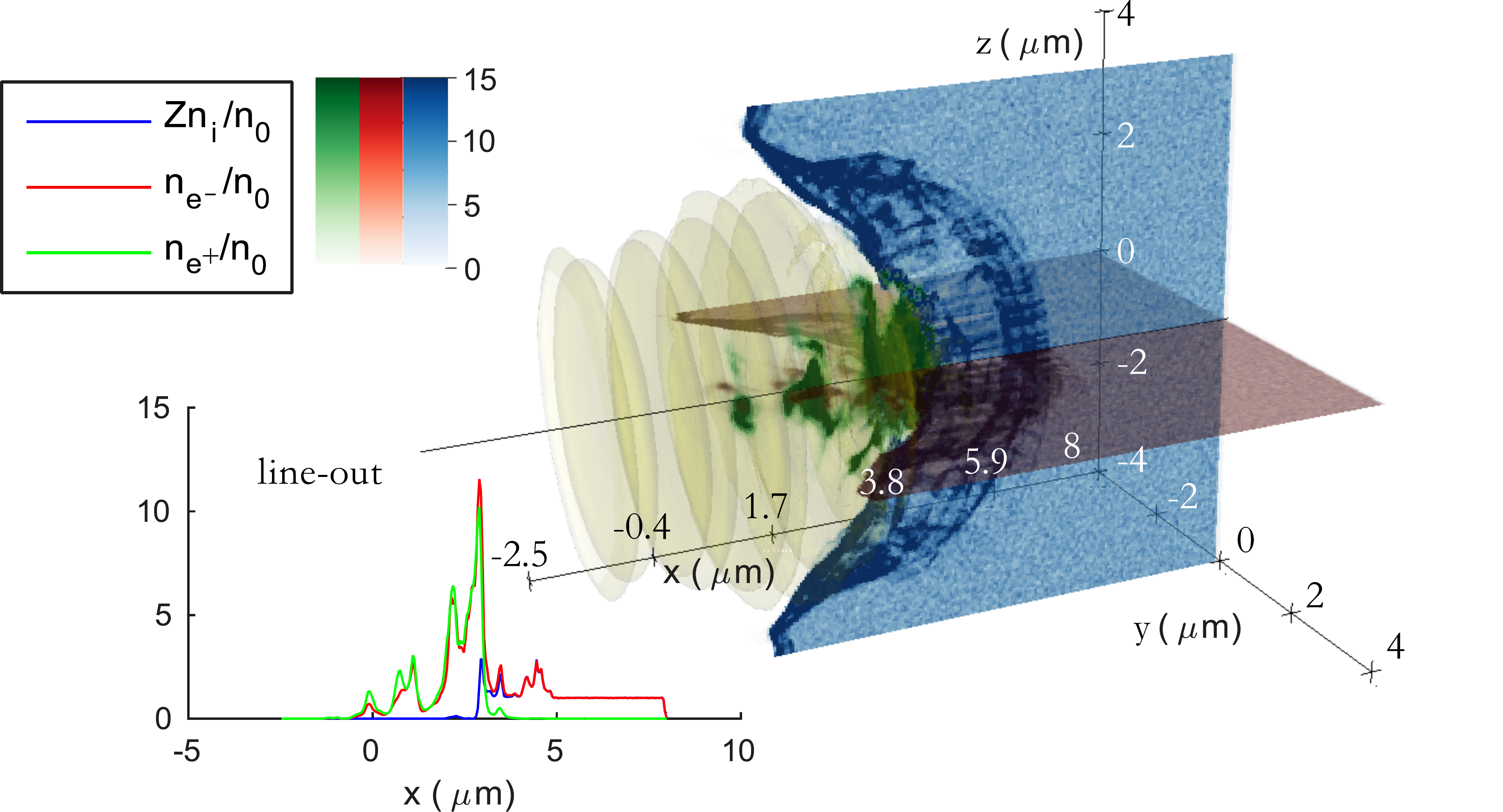}
      \caption{
      QED HB acceleration at $t=6T_{L}$.  The laser is shown in yellow while the 3D target is shown as 2D slices of ions $Zn_{i}/n_{0}$ (blue), electrons $n_{e^{-}}/n_{0}$ (red) and positrons $n_{e^{+}}/n_{0}$ (green), with $Z$ as the ionisation number. Line-outs of particle densities at y=z=0 $\mu$m are also present. They are computed averaging over $12\times12$ cells.}\label{3.pdf}
\end{figure}

 Simulations are performed with the particle-in-cell (PIC) code EPOCH \cite{arber2015contemporary}, which includes both plasma physics and nonlinear QED interactions \cite{furry1951bound,di2012extremely}, the latter according to Ref.~\cite{ridgers2014modelling}. Spin polarization effects \cite{del2017spin} are neglected.
 
Figure \ref{3.pdf} shows 3D simulation results of HB ion acceleration in the regime where QED effects are important, at time $t=6T_{L}$   (where $T_{L}\approx3.33$ fs is the laser period). In this simulation the target is initialized as a $10\;\mu$m $\times8\;\mu$m $\times8\;\mu$m pre-ionized \color{black} aluminum slab with front surface in the plane $x=0$, i.e. the target is much thicker than $\delta_c$ so that the ion acceleration is in the HB regime.  The initial electron density in the target is $n_{0}=10^{24}$ cm$^{-3}$
 and the target is represented by $1.44\times 10^{9}$ macroions and $1.6\times 10^{8}$ macroelectrons (fifth order particle weight functions are used).  \color{black} 
 It is  illuminated by a circularly polarized $1\;\mu\rm m$ wavelength laser of intensity $5\times10^{24}\;\rm W/cm^{2}$. The laser spot spatial profile is a fifth-order supergaussian with $3.3$ $\mu$m full width half maximum and a constant temporal profile, with duration 30 fs. The simulation is performed with $250\times200\times200$ spatial cells corresponding to spatial dimensions of 
 $10.5\;\mu$m $\times8\;\mu$m $\times8\;\mu$m. 
  Absorbing boundaries have been used in the direction of laser propagation and periodic boundaries have been used in directions transverse to this. Doubling the vacuum  space between the target and the laser did not affect the results.  
Convergence testing was performed by doubling the number of macroparticles, spatial gridding and time step. The results have not been significantly affected by this. 
\color{black}

The simulation can be summarized as follows: for $t<3.4T_{L}$, QED effects do not play a significant role in the ion acceleration. As the simulation proceeds, a pair cascade develops, with the number of pairs initially growing exponentially.  After $t\approx3.4T_{L}$, the pair cascade results in the production of an electron-positron pair-plasma with density equal to the relativistically corrected critical density. This pair-plasma grows between the laser and the aluminum ions, forming a pair cushion similar to that described in Ref.~\cite{kirk2013pair}.  This pair cushion absorbs the laser, reducing the energy of the accelerated ions and the efficiency of the acceleration. 
By modifying the absorption, the pair-plasma generated in front of the target reduces both the average ion energy and the efficiency of conversion of laser energy to ion energy. We can determine the energy reduction by analyzing the ratio between the average ion energy in the simulation described above to that in the equivalent simulation where the QED effects are artificially switched off.  At $t=6T_{L}$ this ratio is $ \approx 0.67 $.  The equivalent ratio comparing the efficiency of the ion acceleration, i.e. the total amount of laser energy coupled to the ions, is $\approx0.5$. 

\section{One-dimensional momentum balance model}
\label{section_1D_model}

We will now determine the laser intensities and target densities at which QED effects quench HB ion acceleration by deriving a simple one dimensional model for circularly polarized lasers \cite{macchi2013ion}. We assume that the HB proceeds such that the front surface of the target moves at quasi-constant speed, i.e. ions instantaneously neutralize the charge separation due to electron movement at each time step. In the reference frame of the target's front surface (which we will refer to as the `HB frame'), $<p_{x}>\approx0$ \cite{ridgers2012dense} ($\vec{p}$ is the electron momentum and $x$ as the incident laser direction) due to rapid force balance between the $\vec{v}\times\vec{B}$ force from the laser and the electrostatic force from charge separation. Thus, gamma-ray photons are emitted transversely to the plasma surface and their contribution can be omitted in the HB frame longitudinal, i.e. in the direction of propagation of the laser, momentum balance:
\begin{equation}
\frac{I'}{c}\left(1+R'\right)=2\gamma_{HB}^{2}\rho c^{2} \beta_{HB}^{2}, \nonumber
\end{equation}
$\rho$ is the initial target mass density, $\beta_{HB}=v_{HB}/c$ and $\gamma_{HB}$ are, respectively, the normalized speed and the Lorentz factor of the front surface of the target, i.e.~of the HB frame.  The primed quantities are computed in the HB frame, while, when omitted, quantities are computed in the laboratory frame.  $I'=\left({1-\beta_{HB}})/({1+\beta_{HB}}\right)I$ is therefore the laser intensity in the HB frame and $R'$ is the reflection coefficient in this frame.  $R'$ determines the absorption coefficient $A'$ in the HB frame because transmission is negligible: $A'=1-R'$ \cite{macchi2013ion} \color{black}  
%

From 
the HB frame longitudinal momentum balance, the ion energy in the laboratory frame is \cite{robinson2009relativistically,macchi2013ion,schlegel2009relativistic} 
\begin{equation}
\epsilon=m_{i}c^{2}\frac{2\Pi}{1+2\sqrt{\Pi}}, \quad\quad \Pi=\frac{(1+R')}{2}\frac{I}{\rho c^{3}}.\label{theory en}
\end{equation}
The efficiency of laser conversion into ion energy, $\phi$, is given by 
the ratio between accelerated ion energy per unit of surface and the laser energy per unit of surface, i.e. 

\begin{equation}
\phi=\frac{\epsilon Z n_{0} v_{HB}\tau_{HB}}{I\tau_{P}}. \nonumber
\end{equation}

\noindent Here $\tau_{P}$ is the laser pulse duration and $\tau_{HB}=c\tau_{L}/(c-v_{HB})$ is the interval between when the pulse first strikes the slab and when the trailing edge of the pulse strikes the HB surface. $\phi$ can be written in terms of $\Pi$ as \color{black}
\begin{equation}
\phi=\frac{2\sqrt{\Pi}}{1+2\sqrt{\Pi}}\frac{1+R'}{2}.\label{theory eff}
\end{equation}

QED radiation losses can cause almost complete laser absorption, i.e. $A'\approx1$ and $R'\approx0$. In this case, for strong radiation pressure ion acceleration ($\Pi\gg1$), Eqs.~\eqref{theory en} \& \eqref{theory eff} show that the ion energy is reduced by a maximum of $\sqrt{2}$ and the efficiency of ion acceleration by a maximum of $2$; for weak acceleration ($\Pi\ll1$), i.e.~when ions are non-relativistic (although electrons remains ultrarelativistic), the ion energy and efficiency of acceleration are reduced by factors of $2$ and $2\sqrt{2}$ respectively.  These results are consistent with a similar analysis, limited to the classical radiation reaction force \cite{bashinov2013electrodynamic}.

The scaling laws given in Eqs.~\eqref{theory en} and \eqref{theory eff} require a prediction for the laser absorption caused by QED radiation losses.  Several scaling laws for QED-mediated laser absorption have been discussed in the literature \cite{ji2014energy}, for linear \cite{brady2012laser,levy2016qed} and circular  \cite{capdessus2015influence,nerush2015laser} laser polarization. In the simulation discussed above (and those discussed later) the laser absorption occurs almost entirely in the self-generated pair plasma.  Therefore, QED effects only start to cause significant laser absorption when the density of the pair-plasma generated at the front surface of the target also approaches the relativistically corrected critical density.  This occurs at a time we define as the absorption time $t_{a}$.  As in Ref.~ \cite{grismayer2016laser}, we assume that the absorption in the HB frame is negligible for 
$\tau_{P}<t_{a}$ and is 
\begin{equation}
A'_{HB}=\left(1-\frac{t_{a}}{\tau_{P}}\right), \label{absorption}
\end{equation} for $\tau_{P}>t_{a}$.

It can be shown that, in the laboratory frame, the time required for the pair-plasma to reach the relativistically corrected critical density is  

\begin{equation}
\label{t_a_prediction}
t_{a}=\frac{\gamma_{HB}}{\Gamma(\eta)}\ln\left(\frac{2\gamma_{e}n_{c}}{\gamma_{HB}n_{0}}+1\right), 
\end{equation}
with the electron Lorentz factor $\gamma_{e}$ and the QED parameter $\eta$ (the electric field strength in the electron's rest frame relative to the critical field of QED $E_{crit} = 4.3\times 10^{13}$ G \cite{heisenberg1936folgerungen}) assumed constant in time and computed as described in Eq.~(6) of Ref.~\cite{zhang2015effect},  accounting for the Gaunt factor associated with QED photon emission \color{black}.  $\Gamma(\eta)$ is the rate of the exponential density growth as given in Eq.~(10) of Ref.~\cite{grismayer2015seeded}. In addition we account for the fact that the field inside the target is reduced due to the skin effect in the relativistically overcritical plasma by correcting the intensity as follows: $I_{SD}= I \gamma_{e}n_{c}/(\gamma_{HB}n_{0})$ \cite{ridgers2012dense}. 

\begin{figure}
        \centering
        \includegraphics[width=0.5\columnwidth]{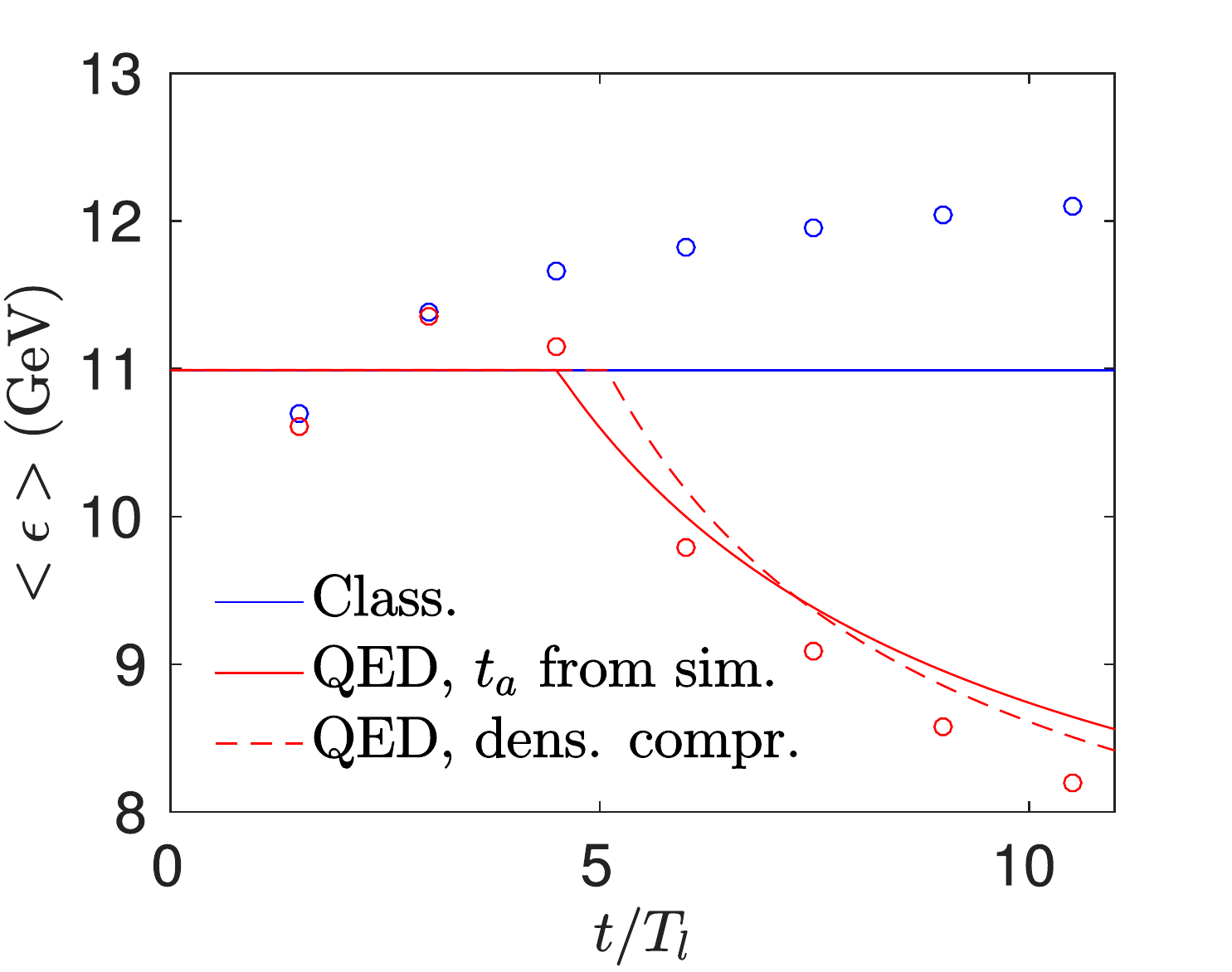}
                        
      \caption{Average ion energy as a function of time, for classical  and QED simulations of a HB acceleration. Dots refers to simulation results while lines to theoretical predictions. In the continuous red line, $t_{a}=15$ fs is provided from simulation, while in the dashed red line electron density increases due to laser compression are considered, according to simulation results.  \label{en_RR.pdf}
}
\end{figure}

The derived expression for $t_{a}$ gives a model which uniquely determines the laser absorption, ion energy and efficiency of ion acceleration for a given laser intensity and target density. To determine the accuracy of this model we compared its prediction to a 1D simulation, with equivalent parameters to the 3D simulation presented above.  10240 cells were used to discretize a domain of 20 $\mu$m, initialized with $1.31072 \times 10^{6}$ macroelectrons and macroions  per cell.  Figure \ref{en_RR.pdf} shows the temporal evolution of the ion energy as predicted by the momentum balance model including QED effects  (two curves labelled `QED, $t_{a}$ from sim.' and `QED, dens. compr.') \color{black} and excluding them (labelled `Class'). These are seen to be in good agreement with the 1D simulations results (shown as dots), where QED effects were included and artificially switched off.  We see that, as predicted by the model develpoed above, after a certain absorption time $t_{a}\approx4.5T_{L}\approx 15$ fs QED effects cause the ion energy to decrease. Figure \ref{en_RR.pdf} shows that ion energy from the model, Eq.~\ref{theory en} agrees well wiith the simulation results if $t_a$ is taken directly from the simulation itself (labelled `QED $t_a$ from sim.'). \color{black}

  Although describing the qualitative behaviour of the ion energy, the model gives, from Eq.~\eqref{t_a_prediction}, ~$t_{a}=7.1$ fs.  The discrepancy with $t_a$ from the simulation is mainly due to the compression of the electron density as the laser strikes the target (which affects the skin layer). In the simulation considered in Fig.~\ref{en_RR.pdf}, this electron density compression is of approximately a factor of two. The result of refining $t_{a}$ to include this density compression is shown in Fig.~\ref{en_RR.pdf}, as the dashed red line (`QED, dens. compr.'). Although improving comparison with simulation, this density compression has been observed to depend on the target density and laser intensity in a complicated way and so a full treatment of it is beyond the scope of this article.  However, on comparison to simulations, the model prediction from Eq.~\eqref{t_a_prediction} has been observed to be accurate enough to predict the general regime of the interaction, i.e. under what conditions QED effects will quench ion acceleration and thus the simple model is useful.  This is discussed further in Sec.~\ref{section_regimes} below. An additional improvement was to delay $t_a$ by one laser period, to account for the time taken to settle to a stationary configuration.  On the contrary, in Sec.~\ref{section_regimes} this did not make any significant difference to the predictions of the model. \color{black}

\color{black}

  In developing the model, we assumed that photons are emitted parallel to the target surface in the HB frame.  This is consistent with the angle of photon emission seen in the 1D simulation $\theta=\arccos(<p_{x}>/<||\vec{p}||>)\approx47^{\circ}$ -- on transformation to the laboratory frame the angle should be $\theta\approx\arccos(\beta_{HB})$ \cite{ridgers2012dense} which gives $\theta\approx46^{\circ}$ for an average ion energy $<\epsilon>\sim10$ GeV.

\section{The two regimes of ultra-high intensity laser-solid interaction: efficient ion acceleration or pair-plasma creation}
\label{section_regimes}

\begin{figure}
        \centering
        \includegraphics[width=.75\columnwidth]{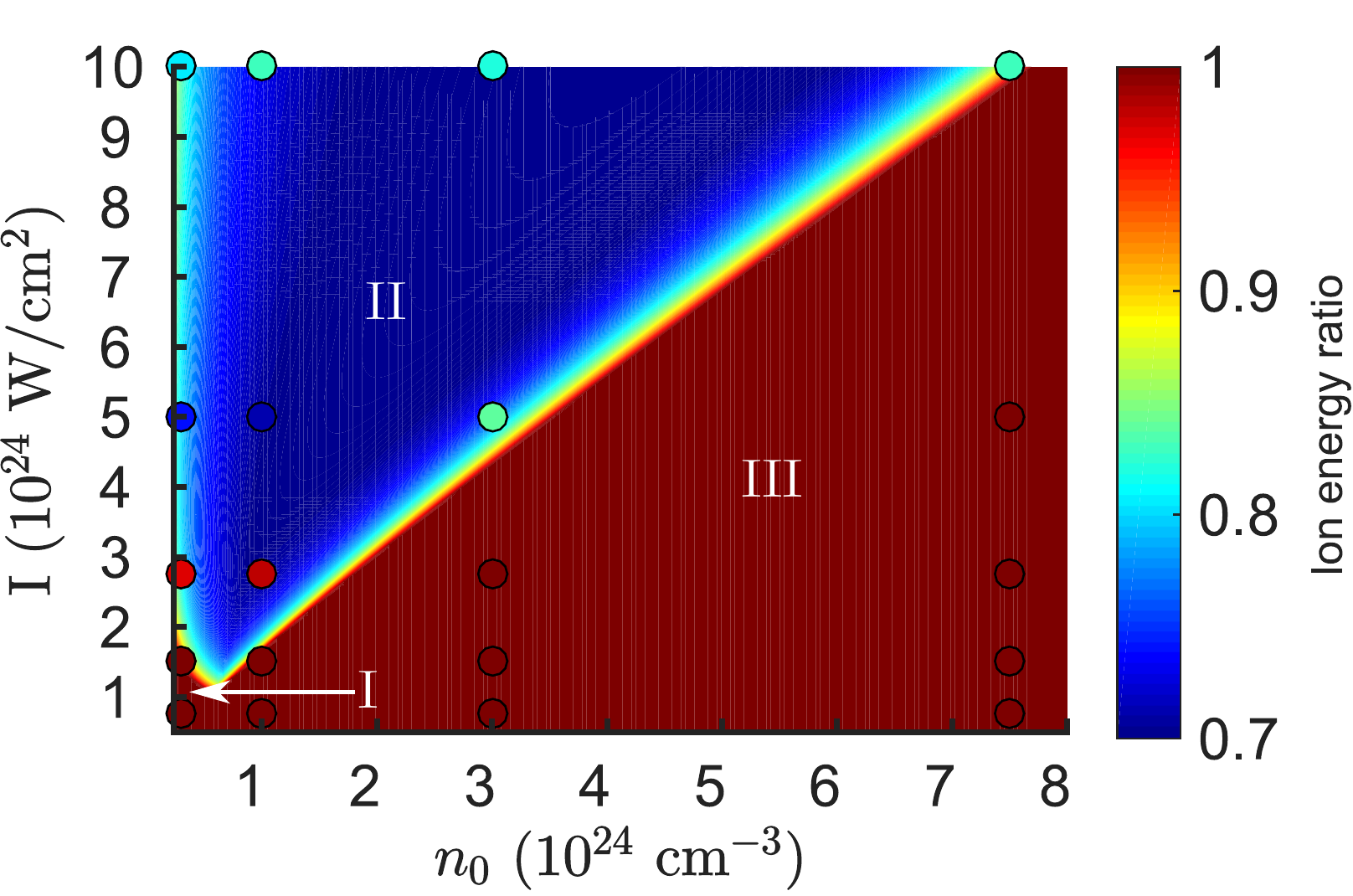}
        \caption{Ratio of QED to classical ion energy , as a function of both laser intensity ($5\times10^{23}$ W/cm$^{2}\le I\le 10^{25}$ W/cm$^{2}$) and initial electron density ($2.3\times10^{23}$ cm$^{-3}\le n_{0}\le 8\times10^{24}$ cm$^{-3}$). Dots represents simulation results, while the colour-plot in the background represents the analytical model. Three regimes are identified: (I) relativistically underdense plasma, (II) QED-plasma regime and (III) relativistically overcritical plasma.  \label{QED_abso.pdf}}
        \end{figure}

Having determined an accurate model for HB ion acceleration including QED effects we can perform a systematic quantitative analysis of QED effects on HB. We limit the simulations used to verify the model to 1D (with identical spatial gridding and number of macroparticles to the 1D simulation described above) and we chose aluminum targets for which QED effects are maximized, according to Eq.~\eqref{theory en} ($\Pi\ll1$).  In Fig.~\ref{QED_abso.pdf} we plot the ratio of the average ion energy including QED effects (by which we mean laser absorption in the self-generated pair-plasma) to that neglecting QED effects, as a function of the initial electron density and of the laser intensity.  The color scale gives the prediction of the model defined above, for a circularly polarized 1 $\mu$m wavelength laser pulse with $\tau_{P}=t=30$ fs. We can identify three distinct regimes. Regime I: here the absorption is negligible and the acceleration can be explained using a classical HB model. This is because the initial target density is too low to initiate a pair cascade (given the probability of pair creation at that laser intensity).  Regime II: as the density increases then a pair cascade can be initiated, resulting in the generation of a critical density pair-plasma and the quenching of ion acceleration. This is coherent with the prediction of efficient gamma-ray emission in near critical plasmas  \cite{brady2013gamma}.  Regime III: if the initial target density is too high then the skin effect screens the laser fields and the cascade does not occur. In regimes I and III laser energy is efficiently (up to 100\%) coupled to ion energy.  In regime II laser energy is efficiently (up to 50\%) coupled to electron-positron pairs and gamma-ray photons and a critical density pair-plasma is generated. As mentioned in Sec.~\ref{section_1D_model}, the absorption time has been delayed by one laser period to account for the time the plasma takes to settle to a stationary state.  However, this did not significantly change the predictions of the model. Figure \ref{QED_abso.pdf} also shows 1D simulation results as colored dots for comparison to the model. The simulations fall into broadly the same regimes as those predicted by the model. 

 Although regime I is characterized as an initially relativistically underdense targets, as the target is illuminated by the laser, the electron compressed to relativistically overcritical densities. Therefore, for densities considered,  regime I can be described by classical HB \cite{robinson2009relativistically}. 

\begin{figure}
        \centering
        \includegraphics[width=0.75\columnwidth]{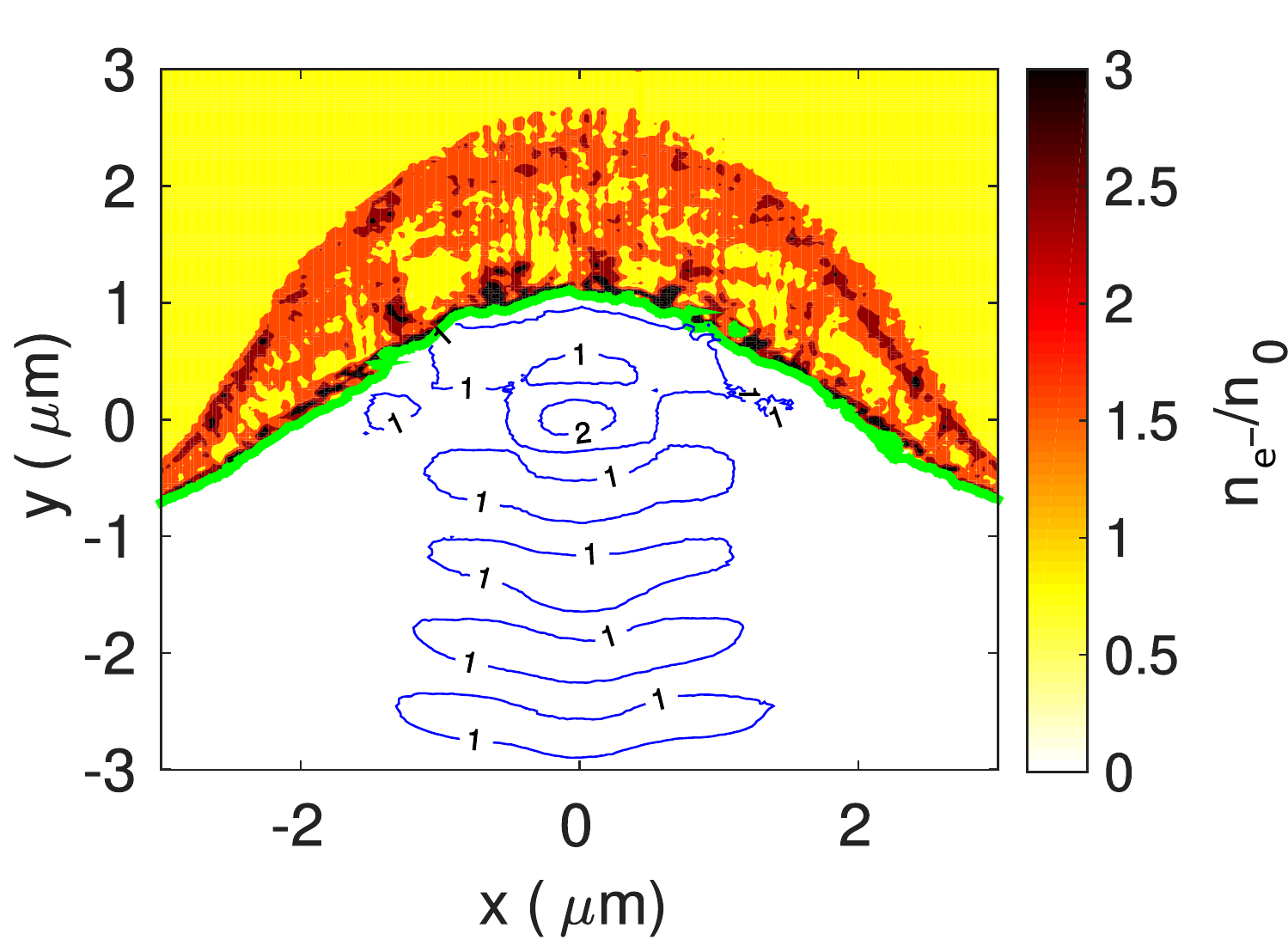}
                        
      \caption{ Electron density and laser field profiles at $t=30$ fs for a 2D simulation of HB acceleration for laser and target parameters in the regime III defined in Sec.~\ref{section_1D_model}. The electron density, normalized to the initial target density, is shown as a red-yellow color plot. The thick green line highlights the relativistically-corrected critical density. The laser electric field magnitude is plotted normalized to the maximum incoming field magnitude as blue isolines.  \label{2d.pdf}
}
\end{figure}

 In Ref.~\cite{may2011mechanism} it has been shown that, in multidimensional simulations, the laser screening due to skin effects in regime III may be prevented by laser modulations of the target surface. This effect has the potential to lead to efficient absorption in regime III, which is not seen in 1D simulations. However, this process is expected to happen on a timescale an order of magnitude longer than those considered. Moreover, in Ref.~\cite{brady2014synchrotron}, the screening has been and resulting low absorption due to QED effects has been seen in simulations with similar parameters to those explored here.  As a test, we have conducted a 2D simulation for initial target and laser parameters in regime III. The simulation is initialized using $600\times600$ spatial grid-cells to describe  a $13\;\mu$m $\times\;16\;\mu$m domain where a $10\;\mu$m $\times\;16\;\mu$m pre-ionized aluminum target of initial electron density $n_{0}=3\times10^{23}$ cm$^{-3}$ is illuminated by a $1$ $\mu$m wavelength laser of peak intensity $2.75\times10^{24}$ W/cm$^{2}$. The laser spot spatial profile is a gaussian (to maximize the effect of the target front-surface deformation) with $3.3$ $\mu$m full width half maximum and a constant temporal profile, with duration 30 fs.  Initially the target is composed by $1.04\times 10^{6}$ macroions and $1.35\times 10^{7}$ macroelectrons, absorbing boundaries have been used in the direction of laser propagation ($x$-axis) and periodic in directions transverse to this. The results, at $t=30$ fs are shown in Fig.~\ref{2d.pdf}: for the timescale considered, the screening is still effective, the laser is reflected at the relativistically-corrected critical density, no significant absorption is seen and no pairs are created.

The three regimes identified in Fig.~\ref{QED_abso.pdf} could be equivalently identified by analyzing Eq.~\eqref{theory eff} instead of Eq.~\eqref{theory en} because they appear as $R'\neq 1$, that is the same in the two cases.\color{black}

\section{The effect of QED processes on light-sail ion acceleration}
\label{section_light_sail}

Another target-parameter which can be varied is the thickness. By decreasing it, we can enter the LS scenario of radiation pressure ion acceleration.  LS ion acceleration is particularly efficient \cite{esirkepov2004highly}: accelerated targets can easily reach the ultrarelativistic limit ($\beta_{LS}\approx 1$). For this reason, the QED parameter $\eta\propto\sqrt{1-\beta_{LS}^{2}}$ \cite{zhang2015effect} is strongly reduced and consequently so are QED effects when compared to the HB cases considered above.   This is supported by 1D simulations similar to those described above, but with the target thickness set by the condition for optimal LS ($\ell=a_{0} n_{c}^{rel}/n_{0}\lambda_{L}/\pi$ where $\lambda_{L}$ is the laser wavelength \cite{macchi2009light}).  In these simulations the target was accelerated much more efficiently than in the equivalent HB simulations (for $I\approx 5\times 10^{24}$  W/cm$^{2}$ and $n_{0}\approx10^{24}$ cm$^{-3}$ we found $\beta_{LS}\approx1$ as compared to $\beta_{HB}\approx 0.7$).  As predicted, in simulations of LS ion acceleration QED effects were indeed negligible for all laser intensities considered here, for example no absorption was seen for the LS simulation with $I\sim10^{24}$ W/cm$^{2}$ compared to 70\% for the equivalent HB simulation. 

\section{Discussion}\label{discussion}

In the interactions of ultra-high intensity laser pulses with matter we can expect particle acceleration over compact, micron-scale, regions \cite{macchi2013ion}, extremely intense bursts of $\gamma$-ray emission \cite{ridgers2012dense} and the creation of dense pair-plasmas \cite{grismayer2015seeded}.  Here we have shown that it is possible to select into which species the laser energy is coupled by the choice of target density and thickness.  By doing so it will be possible to select, in an experiment, which regime one wishes to access: one where ions are predominantly accelerated or one where a dense pair plasma and burst of gamma-rays is generated.  These two regimes were identified in Sec.~\ref{section_regimes} as follows: (i) if the target density is much above or below the relativistically-corrected critical density a cascade of pair production cannot be initiated. In the  first, \color{black} more important case (the density must be sufficient to reflect the laser pulse to efficiently accelerate ions), the electron density in the target is sufficient to effectively screen the laser fields by the skin effect.  This curtails gamma-ray emission from the electrons in the target and thus inhibits the pair cascade.  In this case laser energy is efficiently coupled to ions (in the HB regime of radiation pressure ion acceleration).  (ii) If, on the other hand, the target density is close to the relativistically corrected critical density (as is in fact the case for many solids when the laser intensity is $>10^{23}$\ Wcm$^{-2}$) then the laser fields are not so effectively screened by the target, the electrons feel these fields and initiate a cascade.  In this case a large fraction of the laser energy is absorbed; in principle up to 100\%, although this may place impractical constraints on the target density and laser intensity.  This energy is primarily emitted as gamma-rays from the self-generated pair plasma, which escape the interaction as an ultra-intense (of the same order as the laser-intensity) burst.  A small fraction of these gamma-ray photons are converted to electron-positron pairs, sustaining the pair plasma at the relativistically corrected critical density, which in this case is approximately solid density. 

 Here we have focused on the effect of the produced pair plasma on the ion energy but our model also predicts the efficiency of laser energy converted to ion energy or gamma-rays (the later via the absorption coefficient $A$).  Radiation pressure ion acceleration is in principle a very efficient scheme and even with the maximum predicted efficiency reduction of 65\% it is still relatively efficient, whereas the predicted 50\% reduction in ion energy is very significant for potential applications such as hadron therapy. \color{black}
 
We have also identified another target parameter which could enable the selection of the required regime in experiments: the target thickness.  If the target is sufficiently thin, i.e. less than the relativistic skin depth, then the ion acceleration enters the more efficient LS regime.  Here the target is accelerated to sufficiently high speeds compared to the HB regime that QED effects are negligible (due to the Doppler down shift of the laser intensity in the instantaneous rest frame of the target).  However, the ultimate choice of ion acceleration regime may depend on considerations beyond those investigated here, for example the stability, or lack thereof, of the thin target in the LS regime and pre-pulse effects on the very thin targets required.

In this article we have limited our simulations and scaling laws to the consideration of circularly polarized laser pulses. In reality next generation high intensity lasers will primarily use linear polarization \cite{badziak2004production}.  This is expected to lead to substantially more electron heating than the cases considered here \cite{PhysRevLett.102.095002,PhysRevLett.100.165002}.  This complicates the simple picture we have presented here of the energy partition in the interaction of ultra-high intensity laser pulses with matter, but is not expected to qualitatively change the picture.  The radiation pressure of the pulse is still expected to efficiently accelerate ions 
and pair plasma creation is still expected to be curtailed when the target density exceeds the relativistically-corrected critical density.

Another limitation of the work presented here is the consideration of a very simple target geometry - i.e. a slab.  In reality the high-intensity part of the laser pulse will be preceded by a longer, lower intensity pre-pulse which will pre-heat and so pre-expand the front of the target.  The high-intensity part of the pulse will then have to propagate through a pre-expanded plasma to reach the solid surface of the target.  Instabilities in this plasma could affect the propagation of the pulse.  In simulations of laser-matter interactions at intensities $>10^{23}$\ Wcm$^{-2}$, including a pre-plasma in front of the target, such instabilities have not been observed \cite{badziak2004production}, perhaps due to the increased strength of the ponderomotive force. This leads to significant profile steepening as the plasma is pushed forwards by the pulse, with the target geometry reverting to the simple sharp-edged slab-like profile considered here.  In fact even for the case of a target with a density profile which is initially slab-like, the $\vec{v}\times\vec{B}$ force pushes the electrons into the target (and later pushes electrons and positrons into the pair plasma), leading to a locally enhanced density, as shown in Figs.~\ref{3.pdf} and \ref{2d.pdf}.  This enhancement in the density affects the absorption through the skin effect and is not accounted for in the momentum balance model presented here, limiting the accuracy of this model.  Indeed, in order to obtain accordance between the model described in Sec.~\ref{section_1D_model} and PIC simulations, in Fig.~\ref{en_RR.pdf} it was necessary to fix the absorption time $t_a$ post-hoc from simulation results, while the prediction from Eq.~\eqref{t_a_prediction} was not sufficiently accurate.  However, the prediction of $t_a$ excluding the density enhancement was shown to yield good agreement with PIC simulations for predicting the general regimes of the laser-matter interaction laid out in Sec.~\ref{QED_abso.pdf}.   As demonstrated in Fig.~\ref{en_RR.pdf}, an inclusion of the local enhancement to the electron (and positron) density from simulations would improve the estimate for $t_a$ presented here. 
Due to the observed complicated dependence on laser and target parameters, the compression physics cannot be captured in a simple model. 
 Nevertheless, the current analytical model works sufficiently well to make realistic predictions of the different regimes of multi-PW laser-solid interaction and has the advantage of relative simplicity.  Indeed neglecting the electron density compression for the sake of simplicity is a method previously employed in the description of hole-boring ion acceleration \cite{robinson2009relativistically}. 


\color{black}

\section{Conclusions}\label{Conclusions}

In conclusion, QED effects, specifically the creation of a critical density pair-plasma in front of the target, can quench hole-boring radiation pressure ion acceleration, strongly reducing both the average ion energy (by up to 50\%) and the efficiency of conversion of laser energy to ion energy (by up to 65\%).  We have developed a practical model in order to estimate these reductions.  This model demonstrates the regime where laser energy is efficiently converted to pairs and gamma-rays but also the regimes where this is not the case and laser energy is efficiently converted to ion energy.  We have also found that QED effects do not affect light-sail radiation pressure ion acceleration, when using circular polarization.  These observations will be useful for the design of experiments as they inform the choice of laser and target parameters depending on whether the generation of energetic ion beams or critical density pair-plasmas is the desired aim. Consequently, identifying these regimes of laser-plasma interaction is crucial to the application of next generation lasers as a source of high energy ions, to enable the investigation of dense pair-plasmas in the laboratory, and to produce a very bright source of gamma-rays.

 \section*{Acknowledgments}

This work was funded by the UK Engineering and Physical Sciences Research Council (EP/K504178/1, EP/M018091/1, EP/M018156/1, EP/M018555/1 and EP/P007082/1). 
Computing resources have been provided by STFC Scientific Computing Department's SCARF cluster. The data required to reproduce the results presented here is available at doi:10.15124/65561ffe-049d-4cee-8d79-a8cf590a039e.

%

\end{document}